\documentclass[preprint,showpacs,preprintnumbers,amsmath,amssymb]{revtex4}
\usepackage{graphicx}
\usepackage{bm}

\begin{document}

\title{Full one-loop electroweak corrections to $e^+e^- \to ZH\gamma$ \\ at a Higgs factory}

\author{Ning Liu$^{1,2}$, Jie Ren$^{1,3}$, Lei Wu$^{2}$, Peiwen Wu$^{3}$, Jin Min Yang$^{3}$ \\~ \vspace*{-0.3cm}}
\affiliation{
$^1$ College of Physics $\&$ Electronic Engineering, Henan Normal University, Xinxiang 453007, China\\
$^2$ ARC Centre of Excellence for Particle Physics at the Terascale, School of Physics,
     The University of Sydney, NSW 2006, Australia\\
$^3$ State Key Laboratory of Theoretical Physics, Institute of Theoretical Physics, Academia Sinica,
Beijing 100190, China
 \vspace*{1.5cm} }

\begin{abstract}
Motivated by the future precision test of the Higgs boson at an $e^+e^-$ Higgs factory,
we calculate the production $e^+e^- \to ZH\gamma$ in the Standard Model with complete
next-to-leading order electroweak corrections.
We find that for $\sqrt{s}=240$ (350) GeV the cross section of this production
is  sizably reduced by the electroweak corrections, which is $1.03$ (5.32) fb
at leading order and 0.72 (4.79) fb at next-to-leading order.
The transverse momentum distribution of the photon in the final states is
also presented.
\end{abstract}

\maketitle

\section{ Introduction }

Recently, a Standard Model (SM)-like Higgs boson around 125 GeV was observed by ATLAS and CMS
collaborations at the LHC \cite{atlas,cms}. This discovery is a great step towards the
understanding of electroweak symmetry breaking of the SM. So far, most measurements of the
properties of this new boson are consistent with the SM prediction. The new physics that
affects the Higgs couplings has been cornered to a decoupling region \cite{gardino,belanger}.
Besides, since many extensions of the SM (like the supersymmetric models) contain
a SM-like Higgs boson \cite{higgs-susy} whose properties can be quite similar to the SM Higgs
boson, it is difficult for the LHC to verify whether or not this new boson is the SM one.
In order to precisely study this newly discovered Higgs boson, an $e^+ e^-$ collider,
the so-called Higgs factory, is needed.

In such an $e^+ e^-$ Higgs factory, the properties of Higgs boson can be measured with rather
high precisions \cite{hf1,hf2,peskin}. The dominant Higgs production is the Higgs-strahlung
process $e^+e^- \to ZH$, where the $ZH$ events can be inclusively detected by tagging a
leptonic $Z$ decay without the assumption of the Higgs decay mode.
The individual Higgs decay branching ratios can then be directly measured as the fractions of the
total $e^+e^- \to ZH$ cross section by observing the specific states.
For $\sqrt{s}\sim 240-250$ GeV with an integrated luminosity of 500 fb$^{-1}$, about $O(10^5)$
Higgs bosons can be produced per year, which allows to measure the Higgs couplings at a few
percent \cite{peskin}. So the electroweak radiative corrections should be taken into account
in the theoretical calculations of the production rate. For the process $e^+ e^- \to ZH$, the leading order calculation was performed in
\cite{zh-tree} and the one-loop electroweak corrections were calculated
with the soft-photon approximation in \cite{zh-loop1,zh-loop2,zh-loop3}
(a compact analytical formula for the electromagnetic corrections was given in \cite{zh-loop2}
and a numerical calculation algorithm for the real photon emission was proposed in \cite{zh-real}).

For an $e^+ e^-$ Higgs factory with  $\sqrt{s}\sim 240-250$ GeV another possibly important
process is $ e^+ e^- \to HZ\gamma$. On one hand, it is an important part of the inclusive process
$e^+e^- \to ZH+X$ or can be distinguished for a hard photon; On the other hand, since the $HZ\gamma$
vertex occurs at one loop in the SM, the $HZ\gamma$ couplings is particularly sensitive to possible new physics contributions,
such as the existence of new heavy particles propagating in the loop \cite{peiwen,hancheng}. In this work we calculate this production in the SM with the complete next-to-leading order electroweak (NLO EW) corrections. In Sec. II we will give a description for the analytic calculations.
The numerical results and discussions are given in Sec. III. Finally, we draw our conclusions in Sec. IV.

\section{A description of analytical calculations}
In the SM the process $e^+ e^- \to HZ\gamma$ is induced by the electroweak interaction
at leading order (LO). Due to the small Yukawa couplings, we ignore the contributions
from the Feynman diagrams involving the Yukawa couplings of light fermions.
We denote the four-momenta of initial and final states in the process as
\begin{equation}
  e^+(q_1) + e^-(q_2) \to H(q_3) + Z(q_4) + \gamma(q_5)
\end{equation}
\begin{figure}[t]
  \centering
  \includegraphics[width=15cm]{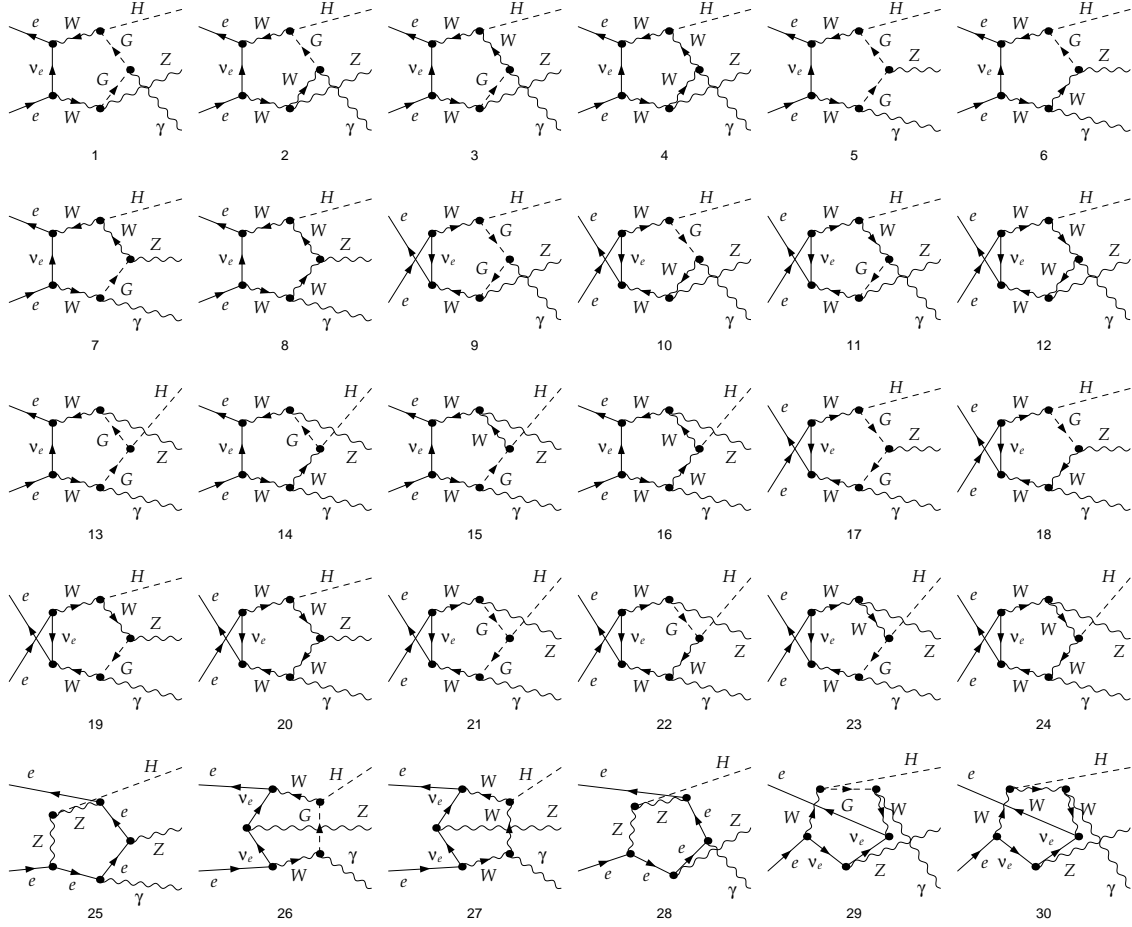}
  \caption{The pentagon diagrams for the process $ e^+ e^- \to HZ\gamma $.}\label{vir}
\end{figure}
The NLO EW corrections ($\Delta\sigma_{EW}$) include two parts:
\begin{itemize}
  \item Virtual correction ($\Delta\sigma_{vir}$).\\
We adopt the dimensional regularization to isolate the ultraviolet divergences (UV)
in the one-loop amplitudes. Then we remove the UV singularities by using the on-mass-shell
renormalization scheme \cite{on-shell}. The pentagon Feynman diagrams in the calculation
are presented in Fig.\ref{vir}. The reductions of N-point ($N\leq4$) tensor integrals are implemented by using the Passarino-Veltman algorithm \cite{loop function}. But for the calculation of the 5-point tensor functions, we adopt the Denner-Dittmaier method developed in Ref.\cite{denner} to reduce the tensor integrals and use our fortran subroutines to perform numerical study, which has been validated in our previous works \cite{bbh,ttz}. We also numerically checked that our results are UV finite.

 \item Real photon radiation ($\Delta\sigma_{real}$).\\
Due to the exchange of virtual photon in the loops, the infrared (IR) divergences can appear
in the virtual correction. According to the Kinoshita-Lee-Nauenberg (KLN) theorem \cite{kln},
these IR divergences will be canceled by the real photon bremsstrahlung corrections in the
soft photon limit. We denote the momenta of initial and final states for the real photon
radiation process as
\begin{equation}
e^+(q_1) + e^-(q_2) \to H(q_3) + Z(q_4) + \gamma(q_5) + \gamma(q_6). \label{re}
\end{equation}
We take the phase-space-slicing method \cite{phase-slice1,phase-slice2} to isolate the
IR singularity in the above process. An arbitrarily small cut-off parameter $\delta_s$
is introduced to split the phase space into soft region ($E_6 \leq \delta_s \sqrt{s} / 2$)
and hard region ($E_6 > \delta_s \sqrt{s} / 2$). So the real photon emission correction
can be decomposed into the soft and hard parts:
\begin{equation}
\Delta \sigma_{real}=\Delta \sigma_{ soft}+\Delta \sigma_{hard}.
\end{equation}
In the soft photon approximation \cite{soft-photon}, we can calculate the soft part of the
correction by using the following equation
\begin{eqnarray}
\label{s} {d} \Delta\sigma_{soft} = d \sigma_{{0}}
\frac{\alpha}{2 \pi^2} \int_{E_{6} \leq \delta_s \sqrt{s} / 2}
\frac{d^3 \vec{q_6}}{2 E_6} \left( \frac{q_1}{q_1 \cdot q_6} -
\frac{q_2}{q_2 \cdot q_6} \right)^2.
\end{eqnarray}
where $E_6 = \sqrt{|\vec{q_6}|^2+m_{\gamma}^{2}}$ and we give a small mass
$m_\gamma$ to the photon to eliminate the IR divergence (we checked that the
dependence on this non-physical mass $m_\gamma$ is exactly canceled
when the real radiation correction and the virtual correction are combined).
Since the hard part of the correction is insensitive to this fictitious photon mass,
it can be directly evaluated by the numerical Monte Carlo method \cite{vegas}.
We notice that there are two photons in the real emission process and one of them should be tagged as the observed hard photon with $p_T > 10$ GeV and $|\eta| < 2$. The phase space integral of these two identical photons in hard part of real emission can be expressed as:
\begin{eqnarray}\label{integral1}
I_{56} &\sim& \frac{1}{2} \left[\int_{E_c}^\infty \frac{d^3 \vec{q_5}}{2 E_5} \int_{\delta_s\sqrt{s}/2}^{E_5} \frac{d^3 \vec{q_6}}{2 E_6} |\mathcal{M}|^2 \theta(E_5-E_6) \right. \nonumber \\
&& \ \left. + \int_{E_c}^\infty \frac{d^3 \vec{q_6}}{2 E_6} \int_{\delta_s\sqrt{s}/2}^{E_6} \frac{d^3 \vec{q_5}}{2 E_5} |\mathcal{M}|^2 \theta(E_6-E_5) \right],
\end{eqnarray}
where the factor $\frac{1}{2}$ is from the identical photons in the final states, and $E_c$ is the energy cut that corresponds to the above hard $p_T$ cut. In order to improve the numerical stability of Eq.(\ref{integral1}), we adopt the method in the Ref.\cite{grace} to carry out the integral Eq.(\ref{integral1}). Since each of the two photons in the final states can be softer or harder than the other one with an equal probability, the Eq.(\ref{integral1}) can be equivalent to:
\begin{equation}\label{integral2}
I_{56} \sim \frac{1}{2}\times2\times \int_{E_c}^\infty \frac{d^3 \vec{q_5}}{2 E_5} \int_{\delta_s\sqrt{s}/2}^{E_5} \frac{d^3 \vec{q_6}}{2 E_6} |\mathcal{M}|^2.
\end{equation}
This means that we can technically assume the photon $\gamma(q_5)$ to be the tagged hard photon and impose a transverse momentum cut $p_T>10$ GeV and pseudo-rapidity cut $|\eta| < 2$ on $\gamma(q_5)$ in the numerical calculations \cite{grace,photoncut}.
\end{itemize}

Finally, the total NLO EW correction of the process $e^{+}e^{-} \to HZ\gamma$ is obtained by
\begin{equation}\label{cs}
\Delta \sigma_{tot} = \Delta\sigma_{vir} + \Delta\sigma_{soft} + \Delta\sigma_{hard}.
\end{equation}
We do the calculations by using the packages FeynArts-3.8 \cite{feynarts},
FormCalc-8.2 \cite{formcalc} and LoopTools-2.8 \cite{looptools} (we have the experience
for using such packages \cite{bbh,ttz,ttr,hwz}). We analytically checked the gauge independence of our LO result by using the Ward identities. We also numerically checked our LO calculations in Feynman gauge(F.G.) with the package \textsf{CompHEP 4.5.2} in unitary gauge(U.G.) \cite{comphep}. In Tab.I, we present the comparison results and find that they are well consistent and are gauge-independent. In addition, by simultaneously interchanging the momenta and the polarization vectors of the two photons, we exploit Bose symmetry of the amplitudes of the real emission processes and found the values of the corresponding amplitudes do not change within numerical precision. We also numerically checked our result for the real radiation correction by
using the Comphep program and found good agreement.

\begin{table}[!h]
  \centering
  \small
  \begin{tabular}{ccc}
     \hline
     $\quad\sqrt{s}\quad$ & $\quad\sigma^{F.G.}_{LO}$ (our)$\quad$ & $\quad\sigma^{U.G.}_{LO}$ (CompHEP)$\quad$ \\
     \hline
     250 & 2.172(2) & 2.172(4) \\
     350 & 5.316(5) & 5.316(9) \\
     500 & 3.562(3) & 3.561(6) \\
     600 & 2.705(3) & 2.705(4) \\
     800 & 1.708(2) & 1.708(3) \\
     1000 & 1.184(1) & 1.184(2) \\
     \hline
   \end{tabular}\caption{The comparison of our LO cross section of $e^+ e^- \to HZ\gamma$ in Feynman gauge with those calculated by \textsf{CompHEP 4.5.2} in unitary gauge.}
\end{table}

\section{ Numerical results and discussions }

\begin{figure}[t]
\centering
\includegraphics[width=15cm,height=7cm]{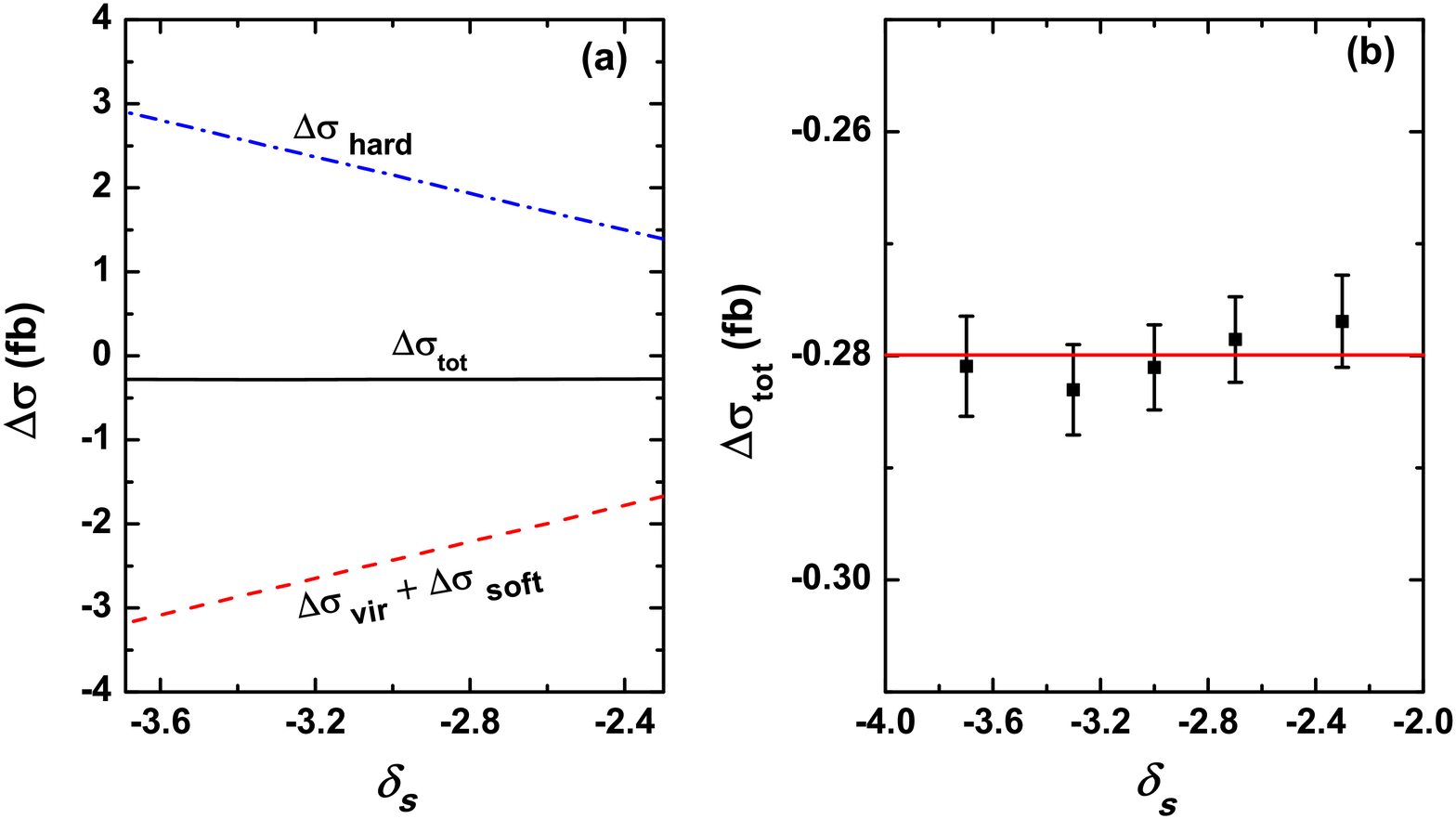}
\vspace*{-0.5cm}
\caption{The one-loop electroweak correction to the cross section
of $ e^+e^- \to HZ\gamma $ versus the soft cutoff $\log\delta_s$ for
$ M_H $ = 125.66 GeV and $ \sqrt{s} $ = 500 GeV: (a) showing respectively
$\Delta\sigma_{hard}$, $\Delta\sigma_{vir}+\Delta\sigma_{soft}$ and
$\Delta \sigma_{tot}$; (b) showing $ \Delta\sigma_{tot} $ with the
calculation errors.}\label{cutoff}
\end{figure}
In the numerical calculations we take the input parameters of the SM as \cite{pdg}
\begin{eqnarray}
m_t=171.2{\rm ~GeV}, ~~m_{e}=0.519991{\rm ~MeV}, ~~m_{Z}=91.19 {\rm
~GeV},\nonumber
\\~~\sin^{2}\theta_W=0.2228, ~~\alpha(m_Z^2)^{-1}=127.918.~~~~~~~~~~~~~~~~
\end{eqnarray}
The Higgs mass is taken as $m_H = 125.66\pm0.34$ GeV \cite{gardino}, which is the combined result
of the measurements of the ATLAS and CMS collaborations.

We numerically check the stability of the results versus the soft photon cutoff parameter in
Fig.\ref{cutoff}, where we assume $\sqrt{s}=500$ GeV and $m_\gamma=10^{-8}{\rm ~GeV}$.
From the left panel of Fig.\ref{cutoff} it can be seen that the values of
$\Delta\sigma_{vir}$, $\Delta\sigma_{hard}$ and $\Delta\sigma_{soft}$ depend on the soft
cutoff $\log\delta_s$, while the total NLO EW correction $\Delta\sigma_{tot}$ is independent
of $\log\delta_s$ within reasonable calculation errors.
Besides, we checked that the total correction is independent of $m_\gamma$ for
a fixed $\delta_s$. Therefore, in the following calculations we take the
$\delta_s=2\times10^{-3}$ and $m_\gamma=10^{-8}{\rm ~GeV}$.

\begin{figure}[t]
  \centering
  \includegraphics[width=15cm]{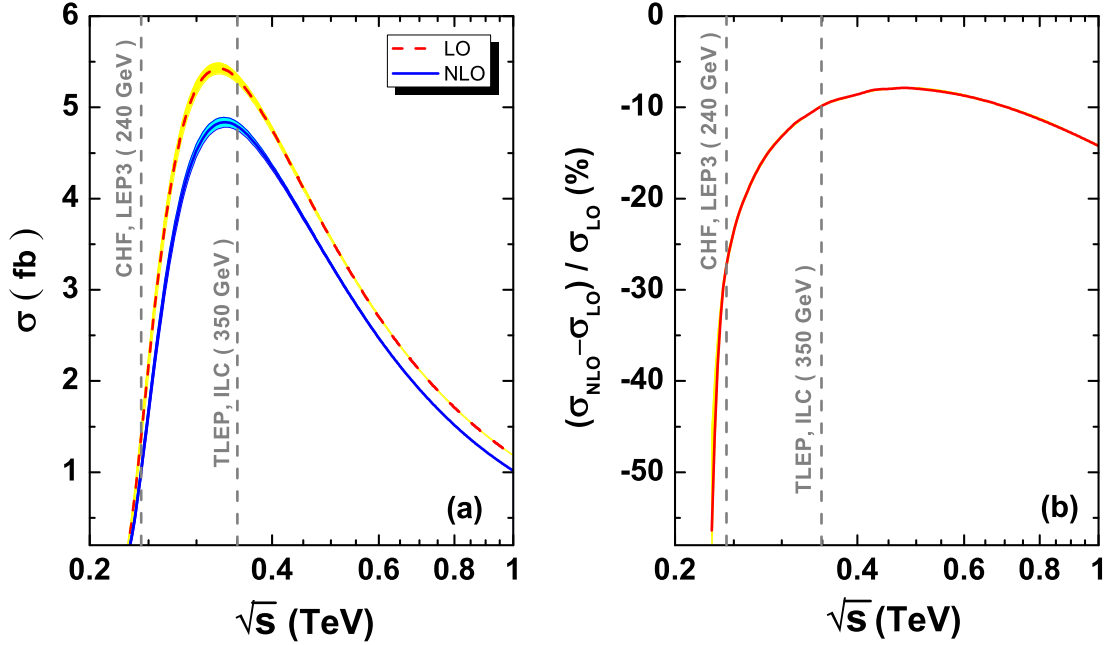}
  \vspace*{-0.5cm}
  \caption{The cross section of $e^{+}e^{-} \to HZ\gamma$ versus $\sqrt{s}$,
showing respectively the LO result and the NLO EW corrections.
The uncertainty caused by the $2\sigma$ range of the Higgs mass (124.98 GeV $<m_H<$ 126.44 GeV)
is also shown (the shaded bands). }
\label{total}
\end{figure}
In Fig.\ref{total} we plot the cross section of $e^{+}e^{-} \to HZ\gamma$
versus the center-of-mass energy $\sqrt{s}$, showing respectively the LO result and the NLO
EW corrections.
We can see that the production rate can reach a few fb in the threshold region
$\sqrt{s} \sim 300-350{\rm ~GeV}$ (maximally it can reach $5.5$ fb at LO and $4.8$ fb at NLO),
and the corresponding EW correction can reach $-12\%$.
For $\sqrt{s} > 400{\rm ~GeV}$, the cross section decreases rapidly due to the suppression
of $1/s$. At a 240 GeV Higgs factory, like the proposed LEP3 or China Higgs Factory (CHF),
the cross section of $e^+e^- \to HZ\gamma$ can reach $1.03$ fb at LO and $0.72$ fb at NLO (the corresponding
EW correction is $-30\%$), while at a 350 GeV Higgs factory, such as the ILC and TLEP,
the cross section of $e^+e^- \to HZ\gamma$ will reach $5.32$ fb at LO and $4.79$ fb at NLO
(the corresponding EW correction is $-9.8\%$).
We can also find that the uncertainty of the cross section caused by
the Higgs mass becomes small with the increase of $\sqrt{s}$.

\begin{figure}[t]
  \centering
  \includegraphics[width=15cm,height=7cm]{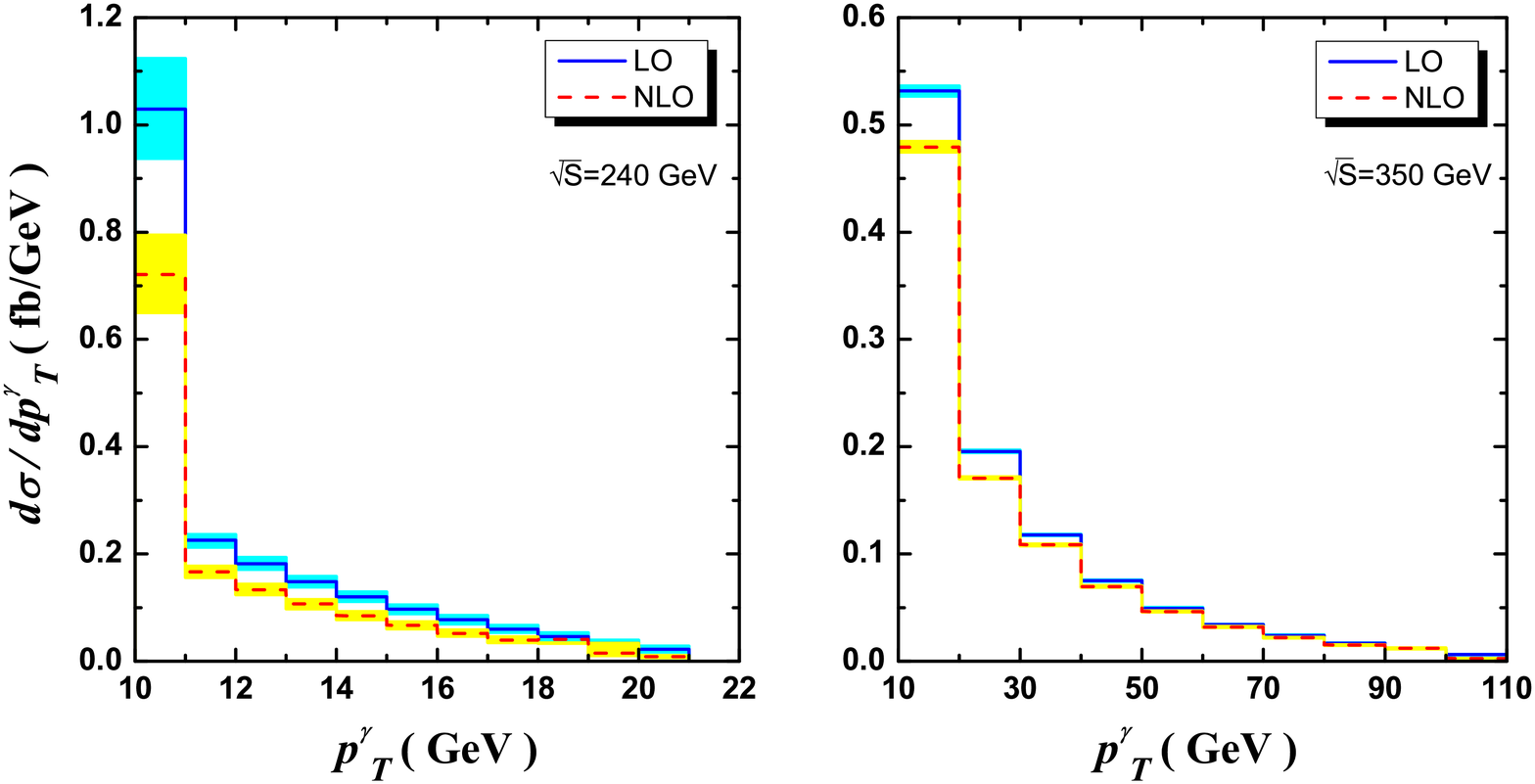}
  \vspace*{-0.5cm}
  \caption{The transverse momentum distribution of the photon at LO and NLO for
the process $e^{+}e^{-} \to HZ\gamma$ with $\sqrt{s}=240,350$ GeV.
The shaded bands correspond to the uncertainty caused by
the $2\sigma$ range of the Higgs mass (124.98 GeV $<m_H<$ 126.44 GeV). }
\label{pt}
\end{figure}

Finally in Fig.\ref{pt} we show the transverse momentum distribution of the photon in the process
$e^+e^- \to HZ\gamma$ at LO and NLO for $\sqrt{s}=240,350{\rm ~GeV}$. It can be seen that the NLO
EW correction can greatly reduce the LO differential cross section at low $p_T$ region.
The impact of the uncertainty of the Higgs mass on the $p_T$ distribution becomes weak as
the collider energy increases. For $\sqrt{s}=240$ GeV most of the events are produced in the
region of $p^{\gamma}_{T}<20{\rm ~GeV}$ due to the center-of-mass energy close to the production
threshold; while for $\sqrt{s}=350$ GeV the $p_T$ value of the photon gets much harder.

\section{Conclusion}
In this work we calculated the cross section of $e^+e^- \to ZH\gamma$ with
complete next-to-leading order electroweak corrections in the SM.
We found that for $\sqrt{s}=240$ (350) GeV the cross section of this production
can reach $1.03$ (5.32) fb at leading order and 0.72 (4.79) fb at next-to-leading
order. In a future $e^+e^-$ Higgs factory, this process can be measured as a precision
test of the SM.

\section*{Acknowledgement}
We appreciate the helpful discussion with Chengcheng Han. Ning Liu would like to
thank Dr Archil Kobakhidze for his warm hospitality in Sydney node of CoEPP in Australia.
This work is supported by the National Natural Science Foundation of China (NNSFC) under
grant Nos.11305049, 11275057, 11275245, 10821504 and 11135003, by Specialized Research Fund for the Doctoral Program of Higher Education
under Grant No.20134104120002, and by the Startup Foundation for Doctors of Henan Normal
University under contract No.11112.

\end{document}